\documentclass[aps,pra,twocolumn,superscriptaddress]{revtex4-1}

\usepackage[utf8]{inputenc}
\usepackage{graphicx}
\usepackage{flafter}

\usepackage{amsmath,amssymb,amstext,amsthm,amsfonts}
\usepackage{braket}
\usepackage{bm}
\usepackage{float}
\usepackage{color}
\usepackage{subfigure}
\usepackage{blindtext}
\usepackage{marginnote}

\begin{document}


\title{Valley filtering in strain-induced  $\alpha$-$\mathcal{T}_3$ quantum dots}

\author{Alexander Filusch}
\affiliation{Institut f\"ur Physik,
Universit{\"a}t Greifswald, 17487 Greifswald, Germany }

\author{Alan R. Bishop}
\affiliation{Science, Technology and Engineering Directorate, Los Alamos National Laboratory, Los Alamos, New Mexico 87545, USA }

\author{Avadh Saxena}
\affiliation{Theoretical Divison, Los Alamos National Laboratory, Los Alamos, New Mexico 87545, USA }

\author{Gerhard Wellein}
\affiliation{Department of Computer Science, Friedrich-Alexander-Universit\"at Erlangen-N\"urnberg, 91058 Erlangen, Germany}

\author{Holger Fehske}
\email{fehske@physik.uni-greifswald.de}
\affiliation{Institut f\"ur Physik,
Universit{\"a}t Greifswald, 17487 Greifswald, Germany }


\date{\today}

\begin{abstract}
We test the valley-filtering capabilities of a quantum dot inscribed by locally straining an $\alpha$-$\mathcal{T}_3$ lattice. Specifically, we  consider an out-of-plane Gaussian bump in the center of a four-terminal configuration and calculate the generated pseudomagnetic field having an opposite direction for electrons originating from different valleys, the resulting valley-polarized currents, and the conductance between the injector and collector situated opposite one another. Depending on the quantum dot's width  and width-to-height ratio, we detect different transport regimes with and without valley filtering for both the  $\alpha$-$\mathcal{T}_3$ and dice  lattice structures. In addition, we analyze the essence of the conductance resonances with a high valley polarization in terms of related (pseudo-) Landau levels, the spatial  distribution of the local density of states, and the local current densities. The observed  local charge and current density patterns reflect the local inversion  symmetry breaking by the strain, besides the global inversion symmetry breaking due to the scaling parameter $\alpha$. By this way we can also filter out different sublattices. 
\end{abstract}

\maketitle

\section{Introduction}
In the emerging field of ``valleytronics," the valley degree of freedom is used to distinguish and designate quantum states of matter.  
For this, the band structure of the system must have at least two inequivalent valleys that take over the role of charge or spin in  
more traditional electronics and spintronics. Two-dimensional condensed-matter materials, such as graphene or semiconducting transition metal dichalcogenides, host an easily accessible electronic valley degree of freedom to encode  information~\cite{RTB07,PXYN07,SYCRRSYX16}. In this respect graphene-based valleytronics seems to be particularly promising. 
This is mostly because of graphene's striking electronic properties~\cite{CGPNG09}, including Dirac-cone functionality which can be tuned by applying external electric fields, even in restricted areas, e.g., by top gates~\cite{GMSHG08}. Another advantage is that diverse graphene nanostructures such as ribbons, rings, quantum dots, or junctions can be manufactured without major problems, whereby  transport through these ``devices" strongly depends on the geometry of the sample and its edge shape~\cite{ZG09}. 

Graphene-based structures also sustain a large amount of strain without breaking because of their strong (planar) covalent $sp^2$ bonds~\cite{LWKH08}. In graphene, the coupling between the mechanical deformation and electronic structure  has remarkable consequences: It introduces an effective gauge field in the low-energy Dirac spectrum. The associated pseudomagnetic field (PMF) has been demonstrated in scanning tunneling microscopy (STM) experiments~\cite{LBMPZGNC10}, which reveal  Landau level (LL) quantization. Most notably, strain-induced PMFs conserve time-reversal symmetry, unlike real magnetic fields,  and therefore point in opposite direction in graphene's  inequivalent valleys  $\mathbf{K}^{}$ and $\mathbf{K}^{\prime}$ related by time-reversal symmetry~\cite{MMP13}. This sign difference together with a spatially varying PMF forms the basis for theoretical proposals to manipulate the valley degree of freedom in graphene-based structures by nanoscale strain engineering~\cite{SFVKS15,CCCFP16,CLFLAN16,SPBJ16,HTWY20}.  In experiments, such local deformation fields can be 
produced and controlled by STM tips~\cite{KKZLW12}. Breaking the valley degeneracy and spatially separating the electrons from different valleys is clearly a prerequisite for every form of valleytronics. In this context, it has been shown that Gaussian bumps lead to different real-space trajectories for $\mathbf{K}^{}$ and $\mathbf{K}^{\prime}$ electrons, and therefore can act  as valley filters and beam splitters~\cite{CFLMS14,SFVKS15,SPBJ16,MP16}.  

The combination of strain, Dirac-cone physics, and flat-band physics in a modified  $\alpha$-$\mathcal{T}_3$ lattice structure is an interesting case to study, not only because the flat band then crosses the nodal Dirac points with peculiar consequences for the Berry phase~\cite{RMFPM14}, Klein tunneling~\cite{SSWX10}, Weiss oscillations~\cite{FD17}, or LL quantization~\cite{FF20}, but also regarding the interplay between the local inversion symmetry breaking by strain and the global one by $\alpha$.  In the  $\alpha$-$\mathcal{T}_3$ structure one of the inequivalent sites of the honeycomb lattice is connected to a site located in the center of the hexagons with strength $\alpha$, i.e., in a certain sense this system interpolates between graphene  $(\alpha=0)$ and dice $(\alpha=1)$ lattices~\cite{Su86}. The dice lattice can be fabricated by growing trilayers of cubic lattices, e.g., SrTiO$_3$/SrIrO$_3$/SrTiO$_3$, in the (111) direction~\cite{WR11}. An $\alpha$-$\mathcal{T}_3$ lattice with an intermediate scaling parameter $\alpha=1/\sqrt{3}$ has been reported for Hg$_{1-x}$Cd$_x$Te at a critical doping~\cite{BUGH09,RMFPM14}. Optical lattice realizations of the $\alpha$-$\mathcal{T}_3$ structure that would allow tuning of $\alpha$ have been also suggested~\cite{BUGH09,RMFPM14}. Based on this background, it  is not surprising that there have been recent activities to exploit valley filtering in the $\alpha$-$\mathcal{T}_3$ and dice  models to realize, for example, a geometric valley-Hall effect~\cite{XHHL17} or magnetic Fabry-P\'erot interferometry~\cite{BMCK20}. Nevertheless the role of nonuniform strain in confined (open) $\alpha$-$\mathcal{T}_3$ nanostructures  is still widely unexplored, especially with regard to the  above-mentioned combination of local  and global inversion symmetry breaking.

In this paper, we address this issue by investigating the transmission of particles through a quantum dot created by an out-of-plane centrosymmetric deformation of an $\alpha$-$\mathcal{T}_3$ lattice in the center of a four-terminal configuration with zigzag terminations. In Sec.~II we introduce our model and discuss the basic impact of the strain-induced PMF with trigonal symmetry in a continuum approach, allowing for an analytical treatment. In particular, taking into account the first-order corrections to the transfer integrals only, we can determine the  (pseudo-) Landau levels (LLs)  and specify their valley dependence with regard to filtering effects.
To also investigate highly strained samples of any geometry and with specific boundaries, we numerically solve the full (tight-binding) lattice-model problem in Sec.~III. For this, we employ the Landauer-B\"uttiker scattering matrix~\cite{Da95} and kernel polynomial~\cite{WWAF06} approaches.  Using the \small{\textsc{kwant}} toolbox~\cite{GWAW14}, we analyze  
the conductance, the valley-polarization, and the local charge and current densities. The results will be discussed with a perspective of potential device applications. Our conclusions are found in Sec.~IV.
\section{Theoretical approach}
\label{theory}
\begin{figure}[htb]
	\centering
	\includegraphics[width=\columnwidth]{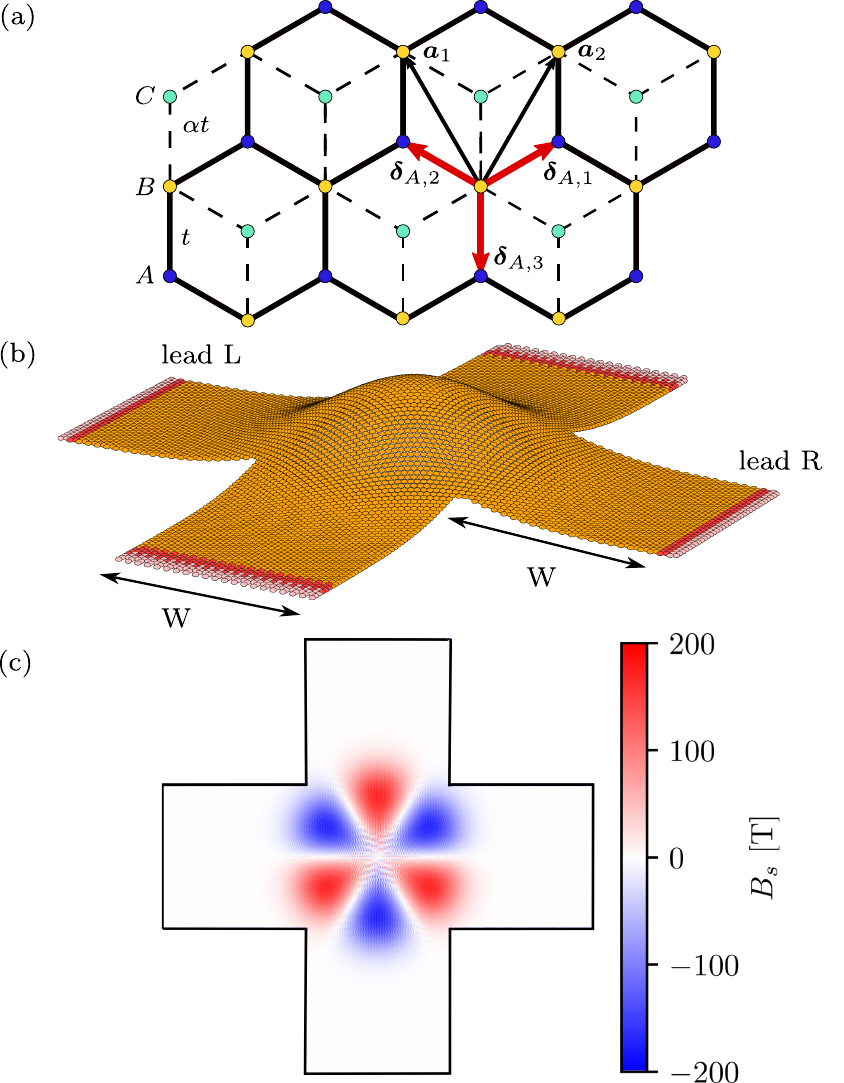}
	\caption{(a) Pristine $\alpha$-$\mathcal{T}_3$ lattice with basis $\{A,B,C\}$  and Bravais lattice vectors $\mathbf{a}_1$ and $\mathbf{a}_2$. Neighboring sites are connected by vectors $\bm{\delta}_{A,j}$ ($j=1,2,3$); the transfer amplitudes on $A$-$B$ and $B$-$C$ bonds are $t$ and $\alpha t$, respectively. 
	(b) Four-terminal configuration with a quantum dot generated by the Gaussian deformation~\eqref{eq:deform}, where $H=17.9$~nm, $\sigma=20$~nm, and $W=50$~nm. 
	(c) Strain-induced pseudomagnetic field calculated for electrons residing in the $\mathbf{K}$ valley. These electrons can pass the quantum dot from L to R whereas electrons stemming from the $\mathbf{K}^\prime$ valley will be reflected.}
	\label{fig1}
\end{figure}

We start from a tight-binding description of the $\alpha$-$\mathcal{T}_3$ lattice by the Hamiltonian 
\begin{align} 
H^\alpha =& - \sum \limits_{\langle i j\rangle}t_{ij} a^\dagger_i b_j^{}-  \alpha \sum\limits_{\langle i j\rangle} t_{ij}  b^\dagger_i c_j^{}\,,\label{eq:Tight-Binding}
\end{align}
where $a^{(\dagger)}$, $b^{(\dagger)}$, and $c^{(\dagger)}$ annihilate (create) an electron in a Wannier state centered at site $A$, $B$, and $C$, respectively. The hopping scaling parameter $\alpha$ interpolates  between the honeycomb graphene lattice ($\alpha=0$)  and the dice lattice ($\alpha=1$) [see Fig.~\ref{fig1}(a)].  In the pristine case, the  transfer amplitude of particles between nearest-neighbor  sites becomes $t_{ij}=t$. Rescaling the energy by $\cos\varphi$, where $\tan\varphi = \alpha$, the Fourier transformed Hamiltonian~\eqref{eq:Tight-Binding}  takes the form
\begin{align}
H^\alpha=\sum \limits_{\mathbf{k}} \psi_\mathbf{k}^\dagger
\begin{pmatrix}
0 &  \cos\varphi f_{\mathbf{k}} & 0 \\
\cos\varphi f^\ast_{\mathbf{k}} & 0 & \sin\varphi  f_{\mathbf{k}}\\
0 & \sin\varphi  f^\ast_{\mathbf{k}} & 0 \\
\end{pmatrix} \psi_\mathbf{k} \label{Tight-Binding-4}
\end{align}
in $\mathbf{k}$ space with $\psi_\mathbf{k}=(a_{\mathbf{k}}, \;b_{\mathbf{k}}, \; c_{\mathbf{k}})$ and 
\begin{align}
f_{\mathbf{k}}=-\sum \limits_{j=1}^3 t_j e^{-i \mathbf{k}\cdot \bm{\delta}_{A,j}^\prime} \label{eq:f_k}\,.
\end{align}

We now consider a lattice distortion by a strain field $\mathbf{u}(x,y)=[u_x, u_y, u_z\equiv h(x,y)]$. Then the displaced lattice coordinates $\mathbf{r}'= \mathbf{r} +\mathbf{u}$ and the bond lengths vary according to $d_{ij}=|\mathbf{r}'_i - \mathbf{r}'_j|$, 
yielding bond-dependent transfer integrals:
\begin{align} 
t_{ij} = t \exp\{-\beta (d_{ij}/a-1)\}\,.  
\label{eq:hopping}
\end{align}
In Eq.~\eqref{eq:hopping}, $\beta=-\partial\log t/\partial\log a$ is the Grüneisen parameter with $a$ being the lattice constant of the unstrained lattice.
These equations will be the basis for the exact numerical study carried out in Sec.~\ref{results}. 
 
At first, however, let us perform some theoretical considerations for an easier interpretation of the results below. If the strain is weak, we only need to take into account first-order corrections to the hopping parameter:
\begin{align}
t_j \simeq  t\left(1- \frac{\beta}{a^2} \Delta_j\right)\,.
\end{align}
Here, $\Delta_j = \bm{\delta}_{A,j} \underline{\epsilon}\, \bm{\delta}_{A,j}$, where the strain tensor  $\underline{\epsilon}$ is given as 
\begin{align}
(\underline{\epsilon})_{ij}& = \partial_i u_j +\partial_j u_i +(\partial_ih)(\partial_j h), & i,j = x,y\,, \label{eq:strain-tensor}
\end{align}
in the framework of continuum theory.  Neglecting other influences of the strain, in the vicinity of the Dirac points $\mathbf{K}^{(\prime)}=\left(\tau \frac{4\pi}{3\sqrt{3}a},0\right)$ with $\tau=1$ ($\tau=-1$), we have 
\begin{align}
f_{\mathbf{K}^{(\prime)}+\mathbf{q}} \simeq &\frac{3at}{2}(\tau  q_x - i q_y) \\\nonumber &+\sum \limits_{j=1}^3 \frac{\beta t}{a^2}  \Delta_j (1 + i \mathbf{q}\cdot \bm{\delta}_{A,j}) e^{i \mathbf{K}^{(\prime)}\cdot\bm{\delta}_{A,j}}.
\end{align}
Inserting this expansion into Eq.~\eqref{Tight-Binding-4}, we find 
\begin{align}
H^\varphi_\tau = \hbar v_\mathrm{F} \mathbf{S}^\varphi_\tau \cdot \left(\mathbf{q} + \frac{ e}{\hbar} \mathbf{A}_\text{s}\right) ,
\label{eq:DW-QT}
\end{align}
with  $v_\mathrm{F}=3at/2\hbar$ (Fermi velocity) and $\mathbf{q}=-i\bm{\nabla}$ (momentum operator in two spatial dimensions). 
The components of the pseudospin vector $\mathbf{S}_\tau^\varphi=(\tau S_x^\varphi, S_y^\varphi)$ in the three-dimensional spin space, 
\begin{align}
S_x^\varphi&=\begin{pmatrix}
0 & \cos\varphi & 0 \\
\cos\varphi & 0 & \sin\varphi \\
0 & \sin\varphi	& 0 
\end{pmatrix},\\[0.1cm]
S_y^\varphi &= \begin{pmatrix}
0 & -i\cos\varphi & 0 \\
i\cos\varphi & 0 & -i\sin\varphi \\
0 & i\sin\varphi	& 0 
\end{pmatrix} \,,
\label{eq:ps-matrices}
\end{align}
represent the sublattice degrees of freedom. Note that the strain-induced vector potential  
\begin{align}
\mathbf{A}_\text{s} = -\tau \frac{\hbar \beta}{2a} \begin{pmatrix}
\epsilon_{xx} - \epsilon_{yy} \\ -2\epsilon_{xy}
\end{pmatrix} \label{eq:As}
\end{align}
depends not only on the two-dimensional strain tensor, but also on the valley index $\tau$ in an explicit way. 
Effectively, it acts as an artificial gauge field that gives rise to a PMF
\begin{align}
B_\text{s} = (\bm{\nabla} \times \mathbf{A}_\text{s})_z = \partial_x A_y - \partial_y A_x  \label{eq:Bs},
\end{align}
perpendicular to the $\alpha$-$\mathcal{T}_3$ lattice plane.

In the presence of a magnetic field, the valley dependence of the (pseudo-) Landau levels (LLs) is of special interest, particularly with regard to valley-filtering effects  when changing the $\alpha$-$\mathcal{T}_3$ lattice scaling parameter $\alpha$ or the direction of the PMF $\gamma=\pm 1$. For this purpose we analyze initially the influence of a constant perpendicular PMF obtained from $\mathbf{A}_\text{s} = -\tau \gamma B_\text{s} y \mathbf{e}_x$ ($B_\text{s}>0$).  Such a PMF can be created by triaxial strain of the lattice~\cite{SPJ16}. Introducing ladder operators $\hat{l}_{\gamma}^{(\dagger)}$ with $[\hat{l}_\gamma, \hat{l}_\gamma^\dagger]=1$, we find for $\gamma=+1$,
\begin{align}
\hat{l}_+^{(\dagger)}=\sqrt{\frac{\hbar}{2 e B_\text{s}}}(q_x \pm i\tau q_y +\tau e B_\text{s}y/\hbar)\,.
\end{align}
For $\gamma=-1$, $\hat{l}_+$ corresponds to $\hat{l}_-^\dagger$. Accordingly, Eq.~\eqref{eq:DW-QT} becomes 
\begin{align}
H^\varphi_{\tau, +} = \tau \hbar \omega_c \begin{pmatrix}
0 & \cos \varphi \hat{l}_+ & 0 \\
\cos\varphi \hat{l}^\dagger_+ & 0 & \sin\varphi \hat{l}_+ \\
0 & \sin\varphi \hat{l}_+^\dagger& 0 
\end{pmatrix} \label{eq:H-psLL}
\end{align}
with $\omega_c = \hbar v_\mathrm{F} \sqrt{2 e B_\text{s}/\hbar}$, which, together with the corresponding expression for $\gamma=-1$, yields the LL spectrum
\begin{align}
E_{\tau, \gamma} = \pm \hbar \omega_c \sqrt{n+\frac{1}{2}(1+\gamma\cos(2\varphi))}\,. \label{eq:psLL}
\end{align}
If we compare this result with the LL spectrum induced by a real magnetic field in the $\alpha$-$\mathcal{T}_3$ lattice [see, e.g., Eq.~(4) in Ref. \onlinecite{RMFPM14}], 
we find that $-\tau$  corresponds to $\gamma$ in Eq.~\eqref{eq:psLL}. This means that the pseudo-LLs are degenerate in valley space.
Of course, there exists an additional zero-energy flat band for $0<\alpha\leq 1$.

Finally, we note that the higher-order contributions (due to larger strain) should give similar corrections as in the case of graphene because $f_\mathbf{k}$ is the same; for a recent review on strain in graphene, see Ref.~\cite{NBOT17}.

\section{Numerical model and results}
\label{results}
In the following calculations we will use `graphene-like' model parameters  $a=0.142$~nm, $t= 2.8$~eV,  and $\beta= 3$~\cite{SFVKS15}, where $t$ sets the energy scale. Furthermore, we consider the four-terminal configuration depicted in Fig.~\ref{fig1} (b)  to study the transport properties of an   $\alpha$-$\mathcal{T}_3$ quantum dot imprinted by straining the lattice with  an out-of-plane  Gaussian bump: 
\begin{align}
h(\rho, \phi)= H \exp{(-\rho^2/\sigma^2)}\,.
\label{eq:deform}
\end{align}
Here, $\rho$ gives the in-plane radial distance from the quantum dot's center. $H$ and $\sigma$ denote 
the magnitude and the characteristic width of the deformation, respectively.

The resulting PMF follows from Eq.~\eqref{eq:Bs} together with Eqs.~\eqref{eq:As} and~\eqref{eq:strain-tensor}:
\begin{align}
B_\text{s} = \tau \frac{4\hbar \beta }{a e}\frac{H^2}{\sigma^3}\left(\frac{\rho}{\sigma}\right)^3 e^{-2(\rho/\sigma)^2} \sin 3\phi 
\end{align}
($\phi$ denotes the polar angle). $B_\text{s}$ is visualized in Fig.~\ref{fig1} (c) in the vicinity of the Dirac point $\mathbf{K}$.  The PMF near  
$\mathbf{K}^\prime$ is simply obtained by reversing the signs. As a result, electrons injected from $\mathbf{K}$ and $\mathbf{K}^\prime$ valleys feel PMFs of opposite 
sign and thus will move in opposite directions. This observation gave rise to the proposal of strain-based valley filtering in graphene \cite{MP16, SPBJ16}.  

\begin{figure}[!]
	\centering
	\includegraphics[width=1\columnwidth]{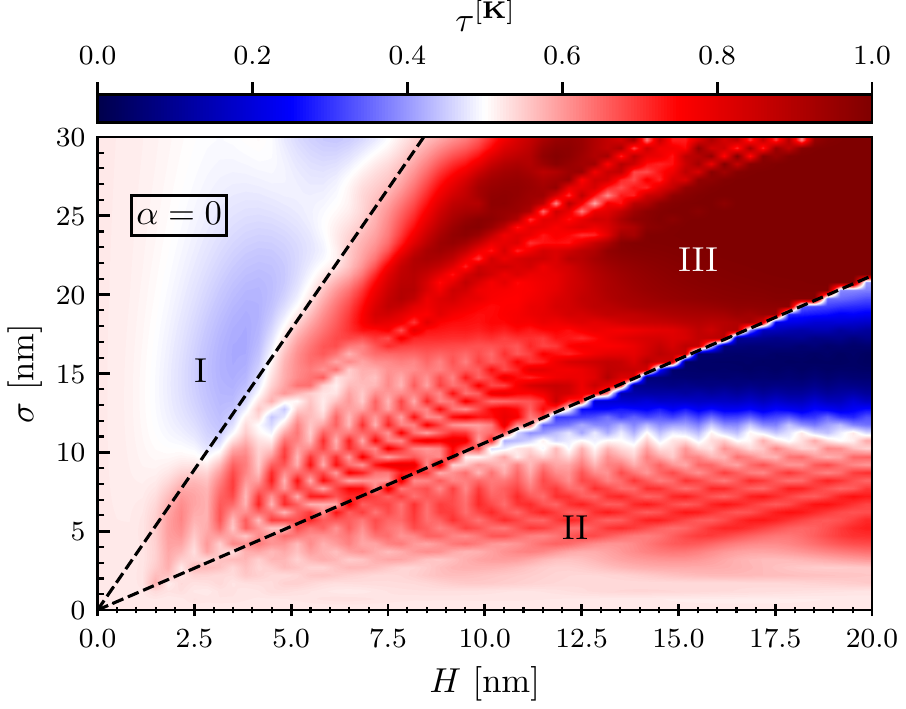}
		\caption{Valley polarization by a strain-induced graphene-based quantum dot.  The contour plot shows $\tau^{[{\mathbf{K}]}}$,  depending  on $H$ and $\sigma$, for the four-terminal configuration in ~Fig.~\ref{fig1}~(b).  The Fermi energy of the injected particles is $E_{\text{F}}\simeq 0.22$~eV.}
	\label{fig2} 
\end{figure}

\begin{figure}[!]
	\centering
	\includegraphics[width=1\columnwidth]{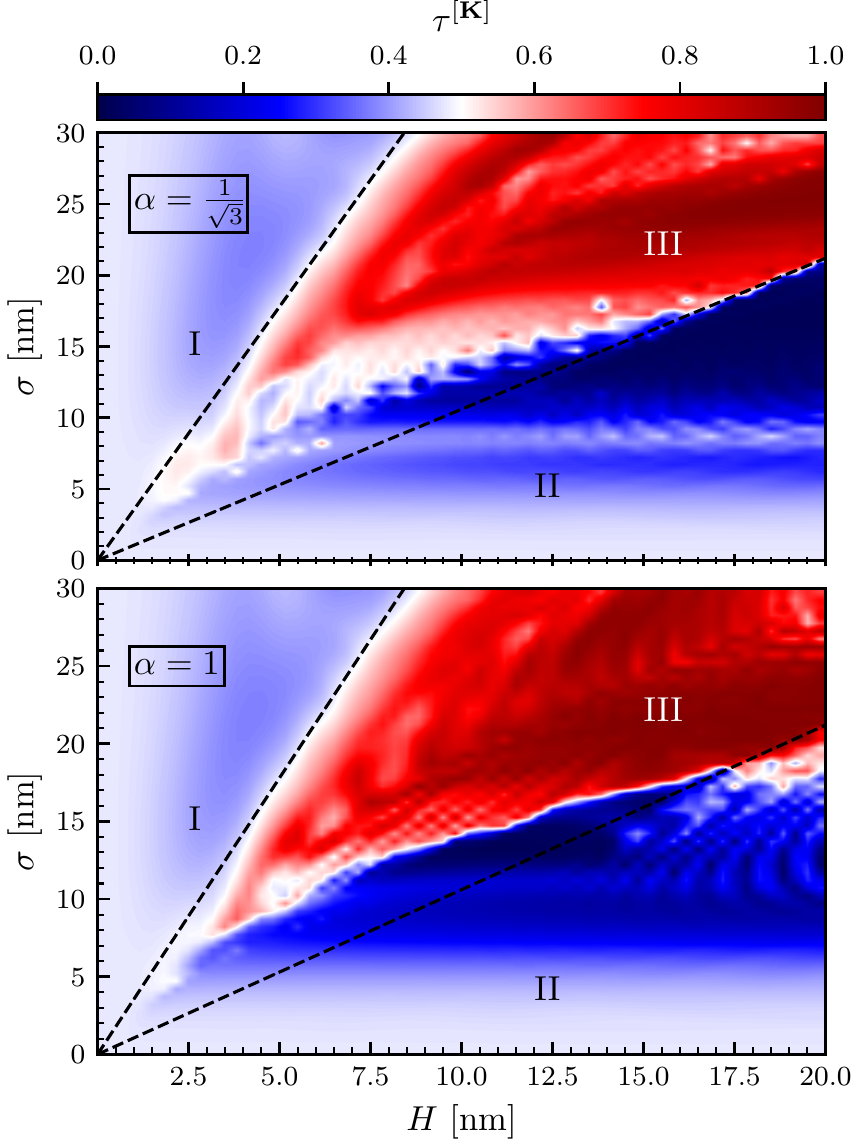}
	\caption{Valley polarization by a strain-induced $\alpha$-$\mathcal{T}_3$ quantum dot.  The contour plots give  $\tau^{[{\mathbf{K}]}}$  as a function of $H$ and $\sigma$ for the four-terminal configurations with $\alpha=1/\sqrt{3}$  (top) and  $\alpha=1$   (bottom), where  $E_{\text{F}}\simeq 0.22$~eV.  We included the linear regime boundaries of Fig.~\ref{fig2} for comparison.}
	\label{fig3} 
\end{figure}
\begin{figure*}[t!]
	\centering
	\includegraphics[width=0.9\textwidth]{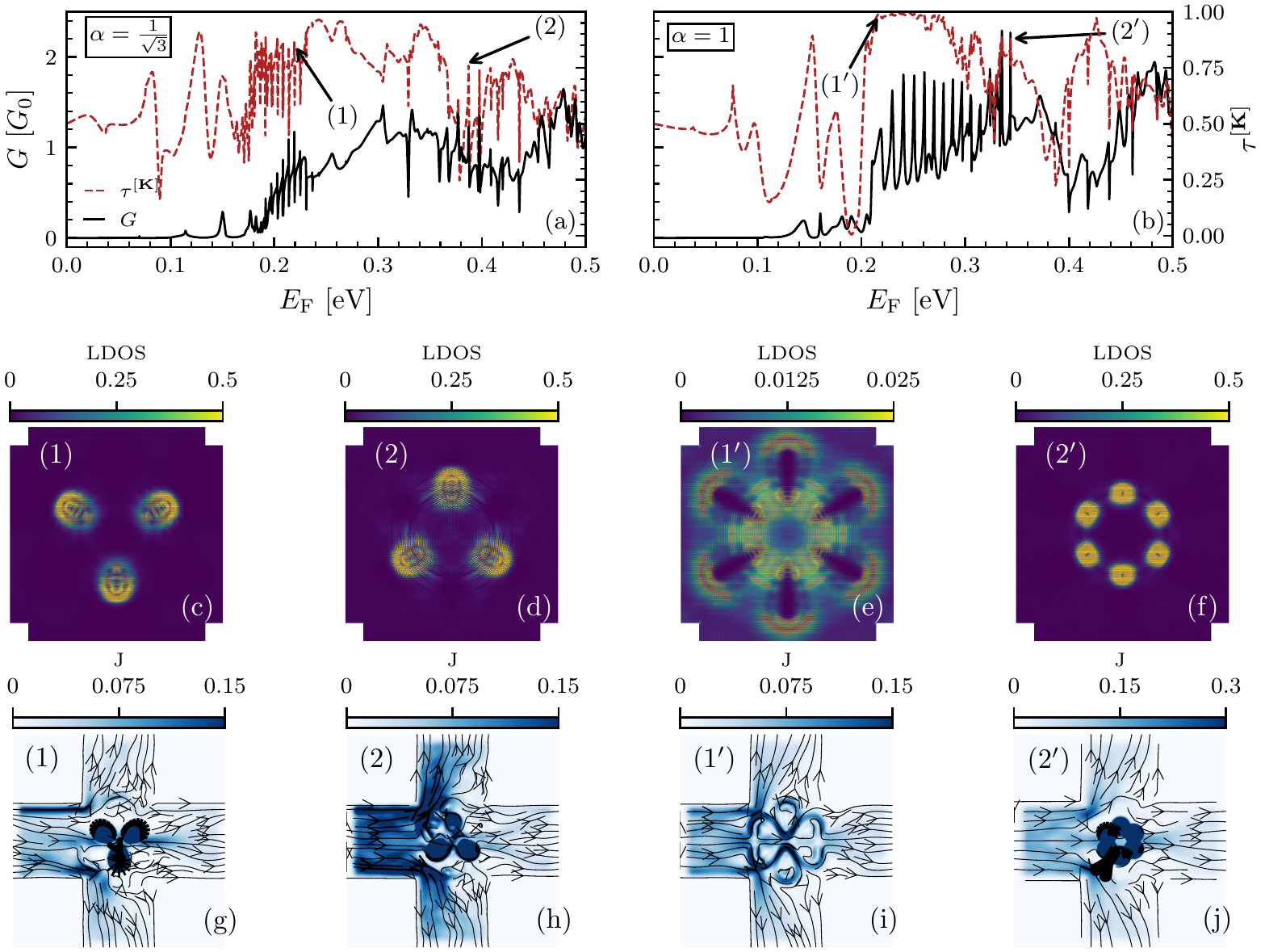}
	\caption{(a) (b) Conductance $G/G_0$ and valley polarization $\tau^{[\mathbf{K}]}$ as a function of $E_\mathrm{F}$ for a strain-induced $\alpha$-$\mathcal{T}_3$  quantum dot with $H=17.9$~nm and $\sigma=20$~nm. For the marked resonances (1) at $E_{\text{F}}=0.219$~eV and (2) at $E_{\text{F}}=0.387$ eV (lattice with $\alpha=1/\sqrt{3}$) we have $\tau^{[\mathbf{K}]}=0.86$ and $\tau^{[\mathbf{K}]}=0.76$, respectively.  For resonances (1$^\prime$) at $E_{\text{F}}=0.219$~eV and (2$^\prime$) at $E_{\text{F}}=0.344$~eV  (dice lattice, $\alpha=1$) we find $\tau^{[\mathbf{K}]}=0.98$ and $\tau^{[\mathbf{K}]}=0.86$, respectively. (c)$-$(f) Zoomed-in LDOS at resonances (1$^{(\prime)}$) and (2$^{(\prime)}$).  (g)$-$(j) Corresponding intensity-coded current densities $|J_{ij}|$ (here, the lines and arrows are a guide to the eye.)}
	\label{fig4}
	\end{figure*}

To determine the conductance $G$ between the left (L) and right (R) leads in the limit of vanishing bias voltage, we use 
the Landauer-B\"uttiker formula~\cite{Da95},
\begin{align}
G = G_0 \sum \limits_{m\in\text{L}, n\in\text{R}} |S_{n,m}|^2\,,
\label{conductance}
\end{align}
where the scattering matrix between all open (i.e., active) lead channels $S_{n,m}$ can be easily calculated with the help of  the \small{\textsc {python}}-based toolbox \small{\textsc{kwant}}~\cite{GWAW14}. In Eq.~\eqref{conductance}, $G_0$ is the maximum conductance per channel. We should also mention the use of zigzag boundaries for the injector (L) and collector (R) leads. In this case the valleys are well separated in momentum space~\cite{RTB07}. The perpendicular leads, which had been added to reduce the leakage of non-valley-polarized currents into the collector~\cite{MP16}, will have armchair boundaries.  This allows us to single out a valley conductance, $G^{[{\mathbf{K}^{(_\prime)}}]}=G_0 \sum_{m\in\text{L}, n\in\text{R}} |S_{n,m}^{[{\mathbf{K}^{(_\prime)}}]}|^2$, which is related to the probability that an injected electron will be transferred in any mode belonging to the $\mathbf{K}$ ($\mathbf{K}^\prime$) valley of the collector. 
Then, with $G = G^{[{\mathbf{K}}]} +G^{[{\mathbf{K^\prime}}]}$, the $\mathbf{K}^{(\prime)}$ valley polarization of the output current can be defined as
\begin{align}
\tau^{[{\mathbf{K}^{(\prime)}}]} = G^{[{\mathbf{K}^{(\prime)}}]} G^{-1} \,.
\end{align}

For validation of our numerical scheme we first reexamined the graphene lattice case ($\alpha=0$) in Fig.~\ref{fig2}, and confirmed the previously found qualitative behavior~\cite{MP16} also for larger values of the Fermi energy. The Fermi energy $E_{\text{F}}=0.219$~eV has been chosen such that it exceeds the barrier produced by the Gaussian bump for a wide range of strain parameters $H$ and $\sigma$. Raising $E_{\text{F}}$  will increase the cyclotron radius and thereby reduce the valley polarization by effectively shrinking the width of the bump.  A maximum valley filtering $\tau^{[{\mathbf{K}}]}$ is observed in regime~III for $\sigma>15$~nm and $1.1 \lesssim \sigma/H \lesssim 3.9$  at $W=50$~nm. In regime~II we have quantum dots with large $H$ and rather small $\sigma$  which generate very high PMFs and therefore tend to repel the electron. This notably weakens the filtering effect. In the blue ``subregime" for $\sigma>10$~nm,  the collector appears to be completely blocked for electrons from the  $\mathbf{K}$ valley~\cite{MP16}. 
The boundary between regimes II and III is almost perfectly linear. The boundary between regimes III and I is more diffuse. The low PMFs in regime~I (due to the small $H$ and large $\sigma$) are clearly inefficient with regard to valley filtering.  

Figure~\ref{fig3} demonstrates that the valley-polarization effect is also observed for $\alpha$-$\mathcal{T}_3$-lattice-based configurations. In the top panel, we have chosen $\alpha=1/\sqrt{3}$ in view of Refs.~\cite{BUGH09,RMFPM14}, whereas $\alpha=1$ in the bottom panel refers to the dice lattice.  Differences compared to the graphene-based system appear, primarily, for  small $H$ and $\sigma$. In particular we find no weak valley-filtering effects in regimes I and II, which in the case of zigzag graphene nanoribbons result from the zero-energy edge state at the $\mathbf{K}$ point, whereas for $\alpha$-$\mathcal{T}_3$ and dice zigzag nanoribbons  the number of zero-energy states at $\mathbf{K}$ is even  due to the additional flat-band state. 

Comparing the valley polarization  in the $\alpha=1/\sqrt{3}$ and $\alpha=1$  lattices, the boundary  between regime~II and III is smeared out in the former case for small to medium $H$. In the dice lattice the most interesting region~III now is more clearly separated from the others, which might be advantageous in terms of possible applications. Note that the ripple structures found in the valley polarization will weaken with increasing size of our configuration (cf. Ref.~\cite{MP16}).

Figures~\ref{fig4}(a) and \ref{fig4}(b) give the conductance $G$ and valley polarization  $\tau^{[{\mathbf{K}]}}$, respectively,  as functions of the Fermi energy $E_{\text{F}}$ for a strain-induced $\alpha$-$\mathcal{T}_3$ quantum dot.  Additional information is provided  by the spatial distribution of the local density of states (LDOS), 
\begin{align}
\text{LDOS}(E)_i = \sum \limits_l |\langle i|l\rangle|^2 \delta(E - E_l)\,,
\label{eq:ldos}
\end{align}
and the local current density,
\begin{align}
J_{ij}^{(m)}= \frac{i}{\hbar}\left[ \langle j |\left(H^\alpha_{ij}\right)^\dagger |i\rangle^{(m)}  - \langle i|H^\alpha_{ij}| j \rangle^{(m)}\right],
\label{eq:current}
\end{align}
where $|i\rangle$ and $|j\rangle$ are the single-particle wave functions projected on the respective sites. Equation \eqref{eq:current} holds for the $m$-th mode injected by the 
lead at energy $E_\mathrm{F}$; the total (incident) current density is $J_{ij}=\sum_m J_{ij}^{(m)}$. Utilizing again \small{\textsc{kwant}}~\cite{GWAW14} and the kernel polynomial method~\cite{WWAF06}, these quantities can be computed very efficiently. The LDOS and $J$ are shown in panels Figs. \ref{fig4}(c)-\ref{fig4}(f) and \ref{fig4}(g)-\ref{fig4}(j), respectively, for the resonances (1), (2), (1$^{\prime}$), and (2$^{\prime}$) marked in panels Figs.~\ref{fig4}(a) and \ref{fig4}(b).

We begin the discussion of how the strained $\alpha$-$\mathcal{T}_3$ quantum dot affects the transport properties of the configuration by examining the conductance (upper panels of Fig.~\ref{fig4}, left ordinate) and the valley polarization (right ordinate).  Of course, a notable current will only flow through the device if the Fermi energy $E_{\text{F}}$  exceeds the barrier produced by the strained quantum dot.  Otherwise the bump, having a high PMF inside, will basically block the flow of electrons towards the collector. We note that our finite quantum system can have a finite, albeit extremely low, transmission probability for ($\mathbf{K}$- or ${\mathbf{K^\prime}}$-valley polarized) electrons with  smaller $E_{\text{F}}$. 
That notwithstanding,  a high valley polarization may occur in this regime just as in the graphene case, where $\tau^{[\mathbf{K}]}$ reaches unity for small energies because of valley--polarized zigzag edge states. 

For $\alpha=1/\sqrt{3}$ above the threshold ($\simeq 0.2$~eV), two regions (bands) in the vicinity of resonance (1) and (2) with a high transmission probability are observed (in the displayed energy interval 0.2$-$0.5~eV). Here, the conductance shows an oscillating behavior that can be attributed to LL quantization (cf. the discussion of Fig.~\ref{fig5} below). In the  dice-lattice case, the conductance features only a single band of resonances between (1$^\prime$) and (2$^\prime$) with particularly high valley polarization and can be similarly attributed to LL quantization. Increasing (decreasing)  the PMF by varying $H$ or $\sigma$ will shift this region to higher (lower) Fermi energies as the LL are proportional to $\sqrt{B_\text{s}}$ [see Eq. \eqref{eq:psLL}]. 

\begin{figure}[t]
	\centering
	\includegraphics[width=1\columnwidth]{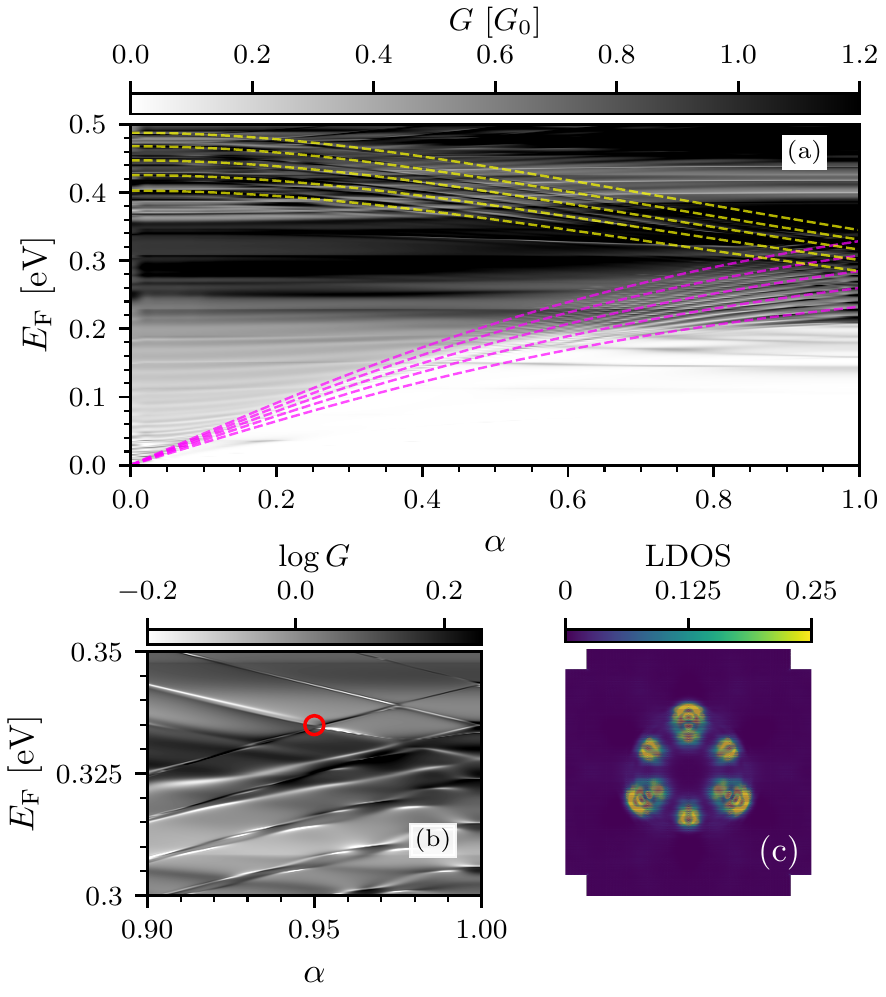}
	\caption{ Transport properties of a strained $\alpha$-$\mathcal{T}_3$ quantum dot  with $A=17.9$~nm, $\sigma=20$~nm. (a) Contour plot of the conductance $G$ in the  $\alpha$-$E_{\text{F}}$ plane. Dashed lines mark the lowest pseudo-LLs, according to  Eq.~\eqref{eq:psLL} with $n=0$, for five equidistant $B_\text{s}$ fields ranging from 100 to 200~T and $\gamma=1$ (yellow), $-1$ (magenta). (b) Zoom-in of the contour plot. (c) LDOS at $E=0.335$ eV and $\alpha=0.95$ [marked in (b) by the red circle].}
	\label{fig5}
\end{figure}

To further characterize the different transport channels we now examine the spatial variation of the LDOS in the quantum dot region. For a graphene-based ($\alpha=0$) quantum dot configuration, the LDOS will show a ``flower"-like pattern with sixfold symmetry, where consecutive `petals' belong 
to the $A$ or $B$ sublattice~\cite{CFLMS14,SFVKS15}. 

In the $\alpha$-$\mathcal{T}_3$ lattice with $\alpha =1/\sqrt{3}$ the inversion symmetry is broken in two ways: Besides the local inversion symmetry breaking by the strain, $\alpha$ itself breaks the inversion symmetry between sublattices $A$ and $C$ on a global scale. As a result, the ``petals" of graphene's ``flower"-like LDOS pattern will have alternating large and small amplitudes [see Figs.~\ref{fig4}(c) and~\ref{fig4}(d)].  The resulting $\nabla$- and  $\Delta$-shaped LDOS patterns with highly occupied sites belong to $A$-$B$ and $B$-$C$ sublattices and correspond to the typical resonances (1) and (2)  in regions with high conductance, respectively.  Moreover, the $\nabla$ ($\Delta$)  LDOS pattern is related to a negative (positive) sign of the PMF.  It is worth noting, however, that both configurations, albeit separated by a large energy gap, possess an almost equally high valley polarization. 

In the $\alpha=1$ dice lattice, we rediscover the sixfold symmetry of the LDOS pattern in view of the threefold symmetries of the positive, respectively negative, strain-induced PMFs [cf. Fig. \ref{fig1} (c)]. This is illustrated by Figs.~\ref{fig4}(e) and~\ref{fig4}(f) for the resonances (1$^\prime$) and (2$^\prime$), respectively, but holds for the whole series of conductance peaks of Fig.~\ref{fig4}(b). Clearly the spatial size and magnitude of the LDOS ``petals" depend on the value of Fermi energy. In this regard, Figs.~\ref{fig4}(e) and~(f) represent rather extreme situations.

Figures~\ref{fig4}(g)$-$\ref{fig4}(j) display the course of the local current density, where its magnitude is coded by the blue-intensity map and the arrows visualize the direction of the  electron flow.  First, it becomes apparent that the electron flow from L to R  is significantly blocked by the bump  and the additional (perpendicular) contacts collect the nonpolarized current very effectively.  Moreover, parts of the electrons are ``confined" in the quantum dot region in long-living resonant states, just as observed for circular graphene quantum dots~\cite{CPP07,HBF13a,PHF13,AU14,FHP15}. This becomes particularly obvious if one looks at Fig.~\ref{fig4}(i), where the current is encircling the PMF.   Nevertheless,  in all cases, substantial amounts of electrons are able to penetrate through the quantum dot  (preferably along the zero-PMF lines) and finally reach the collector R. Recalling the valley polarization according to Figs.~\ref{fig4}(a) and \ref{fig4}(b), we can conclude that this particle stream is made up of electrons belonging to the $\mathbf{K}$ valley. Apparently the current intensities in Figs.~\ref{fig4}(g) and \ref{fig4}(h) nicely feature the $\nabla$ and $\Delta$ $\alpha$-$\mathcal{T}_3$-lattice LDOS patterns in Figs.~\ref{fig4}(c) and \ref{fig4}(d), respectively, and what is more, the current is valley polarized although it seems that the electrons do not feel the full PMF of Fig.~\ref{fig1}(c).

Figure~\ref{fig5}(a) provides a contour plot of the conductance dependence on the scaling parameter $\alpha$ and the Fermi energy $E_{\text{F}}$. In order to assign the onset of the conductance and some of resonances we included the pseudo-LLs~\eqref{eq:psLL} for $n=0$ and $\gamma=\pm 1$ at different PMFs $B_\text{s}$ in the range 100$-$200~T, where the magenta (yellow) curves belong to $\gamma = -1$ (+1). Then, at $\alpha=1/\sqrt{3}\simeq 0.577$, the resonances (1) and (2) from {Fig.~\ref{fig4}(a) fall within the range of the LLs with $n=0$, $\gamma = -1$ and  $n=0$, $\gamma = 1$, respectively, exhibiting  the $\nabla$ and $\Delta$  pattern.
Bearing in mind that $\gamma$ in the PMF takes the role of $-\tau$ for a real magnetic field, the change in the sublattice polarization 
of the LDOS is understandable. This means that exchanging the valleys accounts for the change in the sublattices $A \rightarrow C$. 

In Fig.~\ref{fig5}(b) the region close to the dice-lattice case is enlarged, where the crossing of the $\gamma=\pm 1$ resonances takes place in Fig.~\ref{fig5}(a). Here, the LDOS exhibits overlapping $\nabla$ and $\Delta$ patterns. This is exemplarily demonstrated in Fig.~\ref{fig5}(c). Note that the LDOS shows a similar behavior at the other  (pseudo-LL) ``crossing points" in Figs.~\ref{fig5}(a) and \ref{fig5}(b).

\section{Summary}
To conclude, we have demonstrated how nanoscale strain engineering of pseudomagnetic fields can be used to cause and control valley-polarized transport through an $\alpha$-$\mathcal{T}_3$ quantum dot embedded in a four-terminal configuration with zigzag edges. The strain (pseudomagnetic field) locally breaks the inversion symmetry of the system. By utilizing the \small{\textsc{kwant}} software package, we presented numerically exact results for quantities that characterize the electronic properties  and functionality of the considered device.  Specifically, we discussed the conductance, the valley-filter efficiency and the spatial charge and current density distributions.  We noticed that the conductance resonances with high valley polarization could be related to the  (pseudo-) Landau levels of the continuum quantum dot model. Thereby the local current densities reveal that transmission of electrons with given, let us say, $\mathbf{K}$-valley polarization  is possible and takes place predominantly along the lines of vanishing  pseudomagnetic field;  at the same time, electrons coming from the  $\mathbf{K^\prime}$-valley will be blocked by the quantum dot, and vice versa. For the dice model, at the set of resonances appearing in the first ``conductance band," the maxima in the local density of states show a sixfold symmetry in real space, just as for the graphene case. Any finite $\alpha<1$, however, gives rise to a (global) sublattice asymmetry and therefore creates an energy gap between states belonging to a local density pattern with threefold  $\nabla$ respectively $\Delta$ symmetry. Compared to a graphene-based configuration, for the  dice and $\alpha$-$\mathcal{T}_3$ lattices, the specific (limited) region in the quantum dot's width-and-height parameter space where the maximum valley-filtering effect appears, is much more clearly separated from that with valley-unpolarized transport. This might be advantageous for potential applications. Furthermore, since  $\alpha\neq 0,1$ globally breaks  the inversion symmetry of the lattice, the use of the proposed configuration as an $A$-$C$-sublattice filter is feasible.  Finally, we note  
that  our results are generic to a class of lattices, which means they are applicable to graphene-like materials but also transition metal dichalcogenides and related materials. This also applies to kagome crystals where elastic strain induces the same pseudomagnetic field near the Dirac points as in the $\alpha$-$\mathcal{T}_3$ lattice~\cite{L20}. Equally important, the discussed valley filter effects should stay intact even for weak interactions or spin-orbit coupling as they primarily induce an energy gap.

\section{Acknowledgments} A.F. thanks C. Wurl for valuable discussions. H.F. and G.W. acknowledge the hospitality at Los Alamos National Laboratory where the project has been initiated. The work at Los Alamos National Laboratory was carried out under the auspices of the U.S. DOE and NNSA under  Contract No. DEAC52-06NA25396 and supported by U.S. DOE (A.R.B. and A.S.).


\bibliographystyle{apsrev4-1}

\end{document}